\documentclass[runningheads]{llncs}
\usepackage{graphicx}
\usepackage{tabularx}
\usepackage{wrapfig}
\usepackage{csvsimple}
\begin{document}
\linespread{1.25}
\title{WIP: Are Builders Using Multi-block MEV (MMEV) Opportunities?}
\titlerunning{Abbreviated paper title}
\author{Johannes Rude Jensen\inst{1,2}\orcidID{0000-0002-7835-6424} \and
Victor von Wachter\inst{1}\orcidID{0000-0003-4275-3660} \and
Omri Ross\inst{1,2}\orcidID{0000-0002-0384-1644}}

\authorrunning{Jensen et al.}
\institute{University of Copenhagen, København, Denmark
\and
eToro Denmark, København, Denmark}
\maketitle % typeset the header of the contribution
\begin{abstract}
'Multi-block' MEV (MMEV) denotes the practice of securing k-consecutive blocks in the attempt at extracting surplus value by manipulating transaction ordering. Following the implementation of proposer/builder separation (PBS) on Ethereum, savvy builders can secure consecutive block space by implementing targeted bidding strategies through relays. To estimate the extent to which this practice might be taking place today, we collect data on all bids submitted by builders through relays in the period from the 15th of September (the merge) 2022 until the 31st of January 2023. We hypothesize that builders might secure consecutive blocks in order to deploy sophisticated MMEV strategies, such as creating artificial momentum in Uniswap pools, by withholding and prioritizing transactions from the mempool. In this talk proposal, we present preliminary and non-conclusive results, indicating the builders employ super-linear bidding strategies to secure consecutive block space. We hypothesize that builders act rationally and increase bids only if this is profitable. With this WIP talk proposal, we hope to stimulate an interesting discussion on the feasibility of sophisticated MMEV strategies at SBC'23, with the aim of collecting feedback from researchers and practitioners working on MEV.

\keywords{MEV \and Ethereum \and Multi-block MEV \and Consensus Layer \and Execution Layer}
\end{abstract}
\section{Introduction}
The concept of Maximum Extractable Value (MEV) refers to the economic value extracted through sequencing transactions within the limited state space in a blockchain database. The MEV orchestrator (searcher/builder) extracts rent by structuring transaction flow and injecting their own transactions in ‘bundles’, which are aggregated in blocks and submitted to the elected block proposer, through a centralized relay. The term ‘maximally’ extractable value (MEV) refers to the total opportunity space for value extraction within a given block and is commonly subdivided by extraction strategies considered ‘neutral’ or ‘toxic’. Neutral MEV typically denotes practices that may result in a price improvement for the end user. Examples include ‘just-in-time liquidity’ (JIT) and back-running AMM trades. ‘Toxic’ MEV, on the other hand, refers to practices that will lead to price degradation for the end-user, such as front-running, sandwiching, or equivalent strategies. 
\cite{ref_article8},\cite{ref_article4}. 

Given the rapid rate of development in the space, the categories are non-exclusive and open-ended. Today, the concept of MEV is most commonly associated with the Ethereum blockchain and its ancillary execution domains. At the time of writing, MEV opportunities have generated 182,456 ETH, with the value distribution amongst competing builders heavily skewered towards a set of leading builders of which the top five entities currently produce the lion's share of slots.\footnote{https://mevboost.pics/}The implementation of Proposer Builder Separation (PBS) on the Ethereum main net, following the transition to proof-of-stake, introduces a new enclave of risk vectors as stakeholders compete to extract rent in the PBS paradigm. While PBS has successfully mitigated the risk of proposer/builder collusion without defaulting to traditional first-in-first-out market designs, the concept introduces novel risks.

As a consequence of the epoch construct, the block proposer and committee roles for each epoch are deterministically calculated and revealed one epoch ahead of time\footnote{https://notes.ethereum.org/@vbuterin/SkeyEI3xv}. Practically, this means that, by the time of the last block in epoch N-1 the proposers elected for the subsequent 64 slots, in epoch N and N+1, are known. This introduces two centralization vectors conducive to the extraction of economic value through ‘multi-block' MEV (MMEV) strategies:
First, as noted by \cite{ref_article9}, the probability of a single stake pool operator proposing k-consecutive blocks within a single epoch is a direct property of stake concentration. At the time of writing, consecutive slots assigned to a single pool are a relatively frequent occurrence, occasionally reaching the order of five or even seven consecutive slot elections within a single epoch. If the pool operator of any sequence of proposers elected for k-consecutive slots colludes with a builder by selling consecutive block space, the result is conducive to MMEV extraction. We refer to this extraction category as \textit{collusive MMEV}, as the explicit collusion between the proposer and builder would violate the premise of PBS. 

Second, as $\sim$90\% of active validator nodes are running a generic implementation of the MEVboost sidecar, the selection of bids through relays is a largely non-discretionary and standardized process. In effect, this means that builders can rely on elected proposers to accept a targeted bidding strategy through a relay if (I) the target proposer is running a known version of the MEVboost software, and (II) the target proposer is subscribing to a relay through which the builder can submit bids. We refer to this category of MMEV as \textit{non-collusive}, since builders are abiding by the logic enshrined in PBS.

It is reasonable to assume that large stake pool operators act in good faith, given their mandate as stewards of the ecosystem and exposure to ETH. We operate with the assumption that explicitly collusive MMEV extraction, would eventually be identified by community members and cause significant reputational damage. Yet, it is by no means certain that builders will abstain from implementing sophisticated 'non-collusive' MMEV strategies. On the contrary, one may assume that builders will act explicitly as rational agents and seek to maximize value extraction, through any means necessary. We approach the research question: \textit{ 'To what extent is MMEV extracted on Ethereum today?'}. 

We approximate the occurrence of multi-block value extraction by analyzing progressively increasing successful bids by single builders for consecutive blocks. We hypothesize that builders would only bid above market for consecutive blocks if the MMEV return profile of consecutive blocks exceeds that of standard, single-block MEV. While the results presented in this talk proposal can only be considered indicative at best, we hope to stimulate a fruitful discussion on the implications of increasingly complex extraction strategies by builders. If invited, we aim to collect feedback from the SBC'23 participants on our planned next steps for the study, with an emphasis on identifying MMEV in execution layer data. 

\section{How Could Builders be Exploiting Multi-block MEV (MMEV) Opportunities?}

The notion of MEV extraction spanning several blocks was first introduced as k-MEV in a seminal paper by \textbf{\cite{ref_article0}}. Here, Babel et al. define a weighted notion of MEV (WMEV) which takes the probability of appending multiple blocks in the proof-of-work paradigm into account. 

As noted above, the task of securing consecutive blocks in the PoS paradigm has become significantly easier for builders since the 'merge', as the proposers elected for the subsequent 64 slots are known by the time of the last block in epoch N-1. As a consequence, the formerly probabilistically elected set of block producers can now be considered deterministic, strictly within the context of epoch N and N+1. While knowledge of the proposer set conveys a significant advantage to a hypothetical builder aiming to extract MMEV, the practice still demands a sophisticated operation.

In contrast to standard MEV, securing MMEV opportunities will require information on the changing proposer set. Because some pool operators only work with pre-approved builder addresses, the builder will need to maintain a list of target proposers to which she can submit blocks through one, or more, of the target relays. At the end of epoch N-1, the builder must then identify a set of k-consecutive slots assigned to proposers in epochs N and N+1. Only if the elected proposer for each consecutive slot required for a given strategy matches the requirements, will the builder be able to deploy her MMEV strategy. 

\begin{figure}[!hb]
\includegraphics[width=\textwidth]{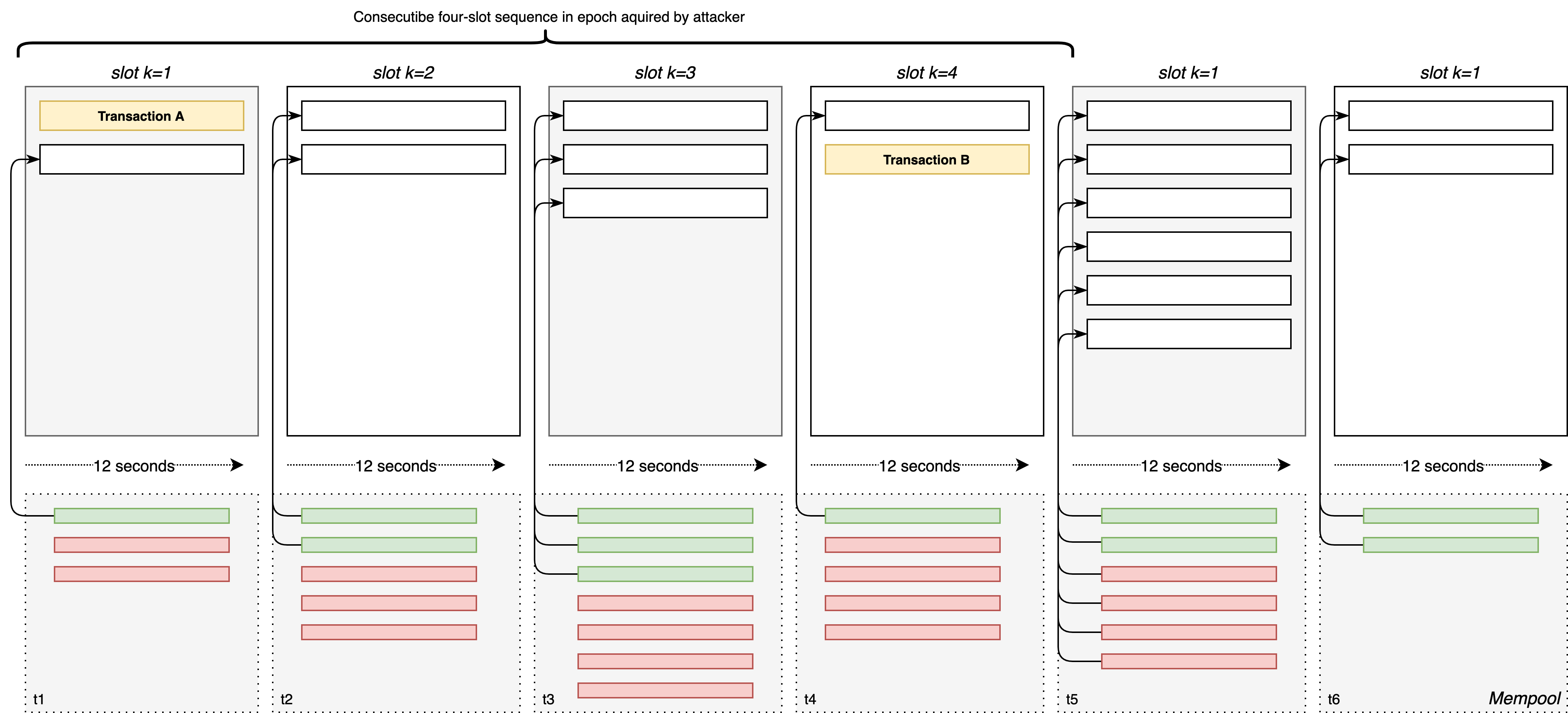}
\caption{Illustration of a hypothetical MMEV momentum strategy. The illustration depicts six consecutive slots, in which the first four consecutive slots have been secured by the antagonistic builder. To capitalize on artificial momentum, the builder first injects a buy transaction in a shallow Uniswap pool, followed by a sequence of organic buy transactions (green) collected from the mempool. To promote price appreciation in the pool, the builder will then withhold all sell transactions in the mempool (red) while collecting as many high slippage-tolerance buy-transactions as possible from the mempool for the duration of the four slots ($\sim$48 sec). At the end of the sequence, the builder can offload their initial buy order and collect the profits generated by the organic buy flow.} A naive builder or proposer will then likely include the withheld sell transactions from the mempool in subsequent blocks. 
\label{fig:mev_example}
\vspace{-15pt}
\end{figure}

For each slot in the target sequence, the elected proposer must (I) have been identified as running the target implementation of the MEVboost software with their proposer client (validator) (II) consume blocks through the target relays, and (III) accept bids from the antagonistic builder. 

The opportunity space for multi-block value extraction is a product of organic on-chain activity and, as such, is in constant flux. Where a standard MEV operation simply aims to identify arbitrage or sandwich opportunities in the mempool, the MMEV orchestrator must operate at a level of sophistication far beyond her peers. We hypothesize that sophisticated builders operating at this level of sophistication might pursue two general 'categories' of MMEV strategies: (I) Discrete strategies, in which builders seek to reach a specific set of pre-defined threshold values provoking an on-chain scenario such as the liquidation of a vulnerable position. For these strategies, the MMEV orchestrator may need to deploy capital throughout the sequence of consecutive blocks or fall short of their goal. (II) Continuous strategies, in which the builder seeks to structure transaction flow throughout the duration of the consecutive slots, by selecting and withholding transactions from the mempool (see Figure \ref{fig:mev_example}). Continuous strategies are likely the favorable option, as they may fail without significant cost to the builder. 

\pagebreak

\section{Data and Methodology}
We aggregate data on all bids submitted to all major relays by builders, since 'the Merge'. The data set comprises the period from block height 15537394 to 16525948 covering the period from the 15th of September 2022 (the 'Merge') to the 31st of January 2023. The dataset contains a total of 994464 slots, averaging 804 bids per slot. The bids are submitted through 132 distinct public keys. We use public sources to match the public keys to 8 builder entities, listed here by volume: Flashbots, Builder0x69, Bloxroute, Beaverbuild.org, Blocknative, Eth-builder.com, x85linux, Eden, and Manifold. While several entities may be controlling the keys, associated with these builders, we assume these agents to collaborate on maximizing MEV returns. Additionally, 7.44\% of the public keys could not be matched to the large builder entities. For these keys, we assume distinct agency. Within the dataset, a total of 92030 ETH was paid by winning bids to the proposers, a value of \$145.68 million, at the time of writing. 

With the intention of producing a superficial test of the general hypothesis, we start the inquiry by deriving the expected number of consecutive blocks through a Monte Carlo simulation. We then compare the results of the simulation with the empirical observations. Throughout, we denote $n$ as the number of slots per epoch ($n$=32), $k$ as consecutive events in a sequence, and $p$ as the probability of success.

\section{Results}
In total, 76.02\% of blocks in the dataset were included by clients implementing the new PBS paradigm. At the time of writing, 88.8\% of blocks are composed using MEVboost. \footnote{https://mevboost.pics}. 
Practically, this means that, in our dataset, a builder can submit a customized block to the elected proposer through a relay in 7.6 out of 10 slots on average.

\subsection{Monte Carlo Simulation}

We start by simulating the expected number of consecutive blocks through a Monte Carlo simulation. In doing so, we randomly sample 10000 instances and calculate the expected value of each event (for 1$\geq$$k$$\geq$10). We model the probability as a Bernoulli experiment, with $p$ being the market share of each builder entity (Table \ref{tab:expected_entity}). Table \ref{tab:results_collections} compares the observed consecutive blocks to the expected consecutive blocks.
\begin{table}
\begin{center}
\caption{\textbf{Expected}. We derive the expected number of consecutive blocks through a Monte Carlo simulation. We randomly sample 10000 instances and calculate the expected value of each event (for 1$\geq$$k$$\geq$10). We model the probability as the Bernoulli experiment with $p$ being the market share for each builder entity.}
\begin{filecontents*}{tables/expected_entity.csv}
builder entity,share,k=1,k=2,k=3,k=4,k=5,k=6,k=7,k=8,k=9,k=10
flashbots,0.2088,131943,26787,5287,1055,202,53,7,1,0,0
builder0x69,0.157,112514,16991,2596,410,56,10,0,0,0,0
bloxroute,0.1114,88166,9496,1001,115,12,1,0,0,0,0
beaverbuild.org,0.1064,85049,8799,897,101,11,1,0,0,0,0
blocknative,0.033,30696,973,19,0,0,0,0,0,0,0
eth-builder.com,0.0312,29026,877,22,0,0,0,0,0,0,0
x85linux,0.0271,25562,698,18,0,0,0,0,0,0,0
eden,0.0074,7342,58,0,0,0,0,0,0,0,0
manifold,0.0035,3378,14,0,0,0,0,0,0,0,0
Total,0.6858,513676,64693,9840,1681,281,65,7,1,0,0
\end{filecontents*}
\csvautotabular{tables/expected_entity.csv}
\label{tab:expected_entity}
\vspace{+20pt}
\caption{\textbf{Observed}. Consecutive blocks by builder entity. 68.58\% of all blocks can be mapped to the largest builders.}
\begin{filecontents*}{tables/leaderboard_entity.csv}
builder entity,blocks,share,k=1,k=2,k=3,k=4,k=5,k=6,k=7,k=8,k=9,k=10
flashbots,207603,0.2088,134724,24898,5319,1194,285,95,26,13,3,2
builder0x69,156094,0.157,100699,18727,4134,916,256,66,21,2,4,0
bloxroute,110822,0.1114,82369,10893,1685,293,67,10,4,1,1,0
beaverbuild.org,105826,0.1064,69033,11596,2737,809,231,78,41,14,8,3
blocknative,32836,0.033,29140,1565,140,21,10,2,0,0,0,0
eth-builder.com,30982,0.0312,25962,2051,249,39,3,0,0,0,0,0
x85linux,26921,0.0271,24295,1147,93,12,1,0,0,0,0,0
eden,7397,0.0074,7293,52,0,0,0,0,0,0,0,0
manifold,3478,0.0035,3427,24,1,0,0,0,0,0,0,0
Total,681959,0.6858,476942,70953,14358,3284,853,251,92,30,16,5
\end{filecontents*}
\csvautotabular{tables/leaderboard_entity.csv}
\label{tab:results_collections}
\end{center}
\end{table}

\pagebreak

\subsection{Mapping Blocks to Builder Entities}
\begin{figure}[!hb]
\centering
\includegraphics[width=\textwidth]{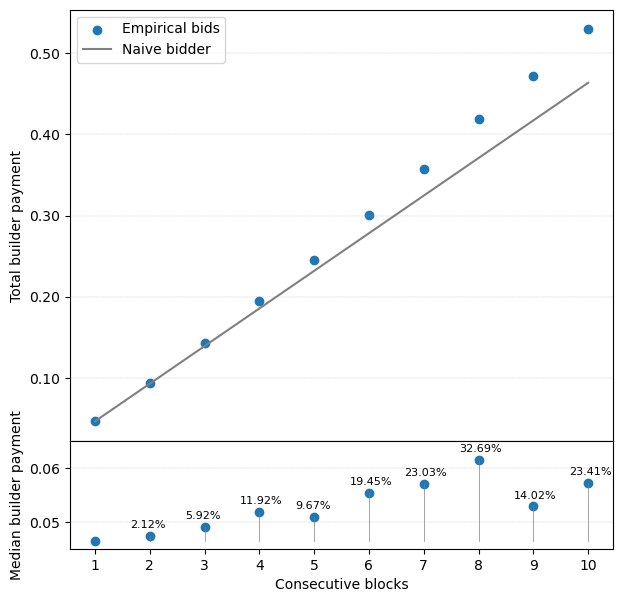}
\vspace{-25pt}
\caption{Builder payments (in ETH) in relation to the number of consecutive blocks by the same builder entity. The lower part depicts the median builder payment: a builder pays 0.046 ETH to secure a single block. Yet, to secure two (three) consecutive blocks, builders are willing to pay 0.047 ETH (0.049 ETH) each, an increase of 2.12\% (5.92\%).} 
\label{fig:payment_entity}

\end{figure}

\begin{figure}[!hp]
\centering
\includegraphics[width=\textwidth]{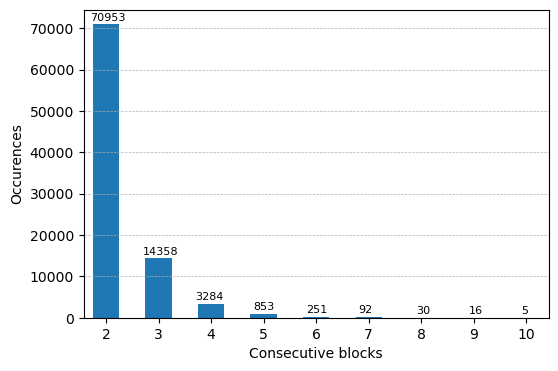}
\vspace{-25pt}
\caption{The figure shows the observed number of $k$ consecutive sequences in the data set. About every 12 seconds, a new block is created. Hence, two consecutive blocks ($k$=2) by the same builder are proposed 514 times a day, on average. We find rare instances of $k$=11, $k$=13, and $k$=17 consecutive blocks in the sample, albeit with a small sample size.}
\label{fig:occurences}

\vspace{+10pt}

\includegraphics[width=\textwidth]{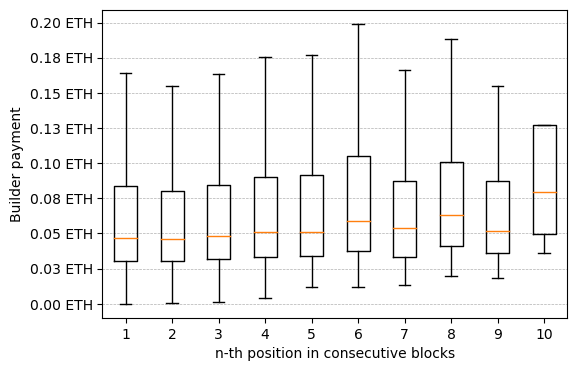}
\vspace{-25pt}
\caption{The illustration shows a boxplot of the builder payment in relation to the n-th position within a sequence of consecutive blocks. The orange line delineates the median bidder payments enclosed by boxes representing the 25\%-quartile and 75\%-quartile. We observe an incremental increase in the average bid value indicating that builders are bidding increasingly higher as the sequence of secured blocks grows.}
\label{fig:payment_position_entity}
\end{figure}

We map the production of blocks to the identified builder entities. As evident, 68.58\% of all blocks can be mapped to the largest builders. Figure \ref{fig:occurences} illustrates the observed number of $k$ consecutive sequences in the data set. About every 12 seconds, a new block is created. Hence, two consecutive blocks ($k$=2) by the same builder are proposed 514 times a day, on average. Notably, we find rare instances of $k$=11, $k$=13, and even $k$=17 consecutive blocks in the sample. Given the small sample size, we exclude these sequences as anomalies. Next, in the upper part of figure \ref{fig:payment_entity} we plot the observed bids by builders securing consecutive blocks.

By plotting the observed bids next to an assumed naive strategy, in which the bidder will always bid at or around the median bid value for the block, we note a super-linear tendency, correlated to the length of the sequence of blocks secured. The lower part in figure \ref{fig:payment_entity} depicts the median builder payment: a builder pays 0.046 ETH to secure a single block.
Figure \ref{fig:payment_position_entity} displays a boxplot of the builder payment in relation to the n-th position within a sequence of consecutive blocks. The median bidder payments are indicated by the orange line. The builder is willing to pay more for the later phases of a sequential sequence.
This suggests that builders are gradually trying to outbid the market for the next slot, extracting the value opportunistically ‘along the way’. 

\section{Discussion and Next Steps}
The preliminary results presented in this talk proposals indicate a trend in which builders employ super-liner bidding strategies to secure consecutive block space. We hypothesize that builders act rationally and increase bids only if this is profitable. Given the dominance of arbitrage and 'sandwich' MEV strategies composed of front and back-running transactions, we hypothesize that savvy builders might secure consecutive block space in order to create artificial trading momentum 

Yet, given the preliminary nature of this work, this talk proposal includes several critical omissions. We hope to have the opportunity to collect feedback from colleagues at SBC'23. First, there may be several reasons why builders would bid above the market to secure consecutive blocks without exploiting MMEV, if not simply by chance. These reasons would all yield false positives in the analysis, based on the assumptions presented here.

Second, the present results do not include any tangible examples of MMEV in the execution layer. Identifying and labeling complex extraction strategies will require collecting execution layer data from the slots featured in the data set and quantifying the prevalence, types, and possibly USD-denominated returns secured through MMEV opportunities. As a consequence of these omissions, the results presented here should be considered non-conclusive and indicative, at best. 

Nevertheless, we hope that the feasibility of MMEV exploitation delineates an interesting area of discussion for researchers and practitioners at SBC'23. As noted by \cite{ref_article3} builders are able to impose many arbitrary rules on a sequence of blocks for which they are ordering transactions. The most dominant expression of builder preferences is, arguably, outright exclusion. 

Fair transaction sequencing in BFT systems presents a fascinating problem, for which there exists a range of solutions. Recent proposals for reducing the implications of MEV on the Ethereum blockchain, range from smart contract designs in the execution layer \cite{ref_article5} to the implementation of new encryption techniques in the consensus layer \footnote{https://ethresear.ch/t/shutterized-beacon-chain/12249}.

\bibliographystyle{splncs04}
\bibliography{mybibliography}

\pagebreak

\end{document}